\documentclass{JHEP3} 

\usepackage{cite,graphicx,multicol}

\newcommand{\s}{\small}
\newcommand{\nn}{\nonumber}
\newcommand{\po}{\phantom{000}}

\newcommand{\IM}{\mbox{\rm Im}}

\newcommand{\eqn}[1]{(\ref{#1})}

\newcommand{\smvs}{\vbox{\vskip 8mm}}

\newcommand{\MSb}{{\overline{\rm MS}}}
\newcommand{\sfrac}[2]{\mbox{$\frac{#1}{#2}$}}
\newcommand\fverb{\setbox\pippobox=\hbox\bgroup\verb}
\newcommand\fverbdo{\egroup\medskip\noindent%
			\fbox{\unhbox\pippobox}\ }
\newcommand\fverbit{\egroup\item[\fbox{\unhbox\pippobox}]}
\newbox\pippobox

\title{Contour-improved versus fixed-order perturbation theory in
       hadronic \boldmath{$\tau$} decays}

\author{Matthias Jamin \\
        Instituci\'o Catalana de Recerca i Estudis Avan\c{c}ats (ICREA) \\
        Theoretical Physics Group, IFAE/UAB,
        E-08193 Bellaterra, Barcelona, Spain \\
        E-mail: \email{jamin@ifae.es}}

\preprint{}
\preprint{UAB-FT-587}

\abstract{
The hadronic decay rate of the $\tau$ lepton serves as one of the most precise
determinations of the QCD coupling $\alpha_s$. The dominant theoretical source
of uncertainty at present resides in the seeming disparity of two approaches
to improving the perturbative expansion with the help of the renormalisation
group, namely fixed-order and contour-improved perturbation theory. In this
work it is demonstrated that in fact both approaches yield compatible results.
However, the fixed-order series is found to oscillate around the
contour-improved result with an oscillation frequency of approximately six
perturbative orders, approaching it until about the 30th order, after which
the expansion reveals its asymptotic nature. Additionally, the renormalisation
scale and scheme dependencies of the perturbative series for the $\tau$
hadronic width are investigated in detail.
}


\begin{document}


\section{Introduction}

Already more than a decade ago it was realised that the hadronic decay of the
$\tau$ lepton could serve as an ideal system to study low-energy QCD under
rather clean conditions \cite{bnp:92,bra:89,bra:88,np:88}. In the following
years, detailed investigations of the $\tau$ hadronic width as well as
invariant mass distributions have served to determine the QCD coupling
$\alpha_s$ to a precision competitive with the current world average
\cite{dhz:05,aleph:05,opal:99,aleph:98,cleo:95,aleph:93}. The experimental
separation of the Cabibbo-allowed decays and Cabibbo-suppressed modes into
strange particles \cite{dhz:05,opal:04,aleph:99} opened a means to also
determine the quark-mixing matrix element $|V_{us}|$
\cite{gjpps:04,jam:03,gjpps:03} as well as the mass of the strange quark
\cite{bck:04,cdghpp:01,dchpp:01,km:00,kkp:00,pp:99,ckp:98,pp:98}, additional
fundamental parameters within the Standard Model.

The starting point for a QCD analysis of the $\tau$ hadronic width
$R_\tau$ is the finite energy sum rule (FESR) \cite{bnp:92,kpt:83}
\begin{equation}
\label{RtauPi}
\frac{\Gamma(\tau\to{\rm hadrons}\;\nu_\tau)}{\Gamma(\tau\to\mu\,
\bar\nu_\mu\nu_\tau)} \,\equiv\, R_\tau^\Pi(s_0) \,=\, \int\limits_0^{s_0}
w_\tau(s) \,\rho_\tau(s)\,ds \,=\, \frac{-1}{2\pi i}\!\!\oint\limits_{|s|=s_0}
\!\! w_\tau(s)\,\Pi_\tau(s)\,ds \,.
\end{equation}
The FESR can easily be derived from Cauchy's theorem and the fact that the
exact, non-perturbative correlation function $\Pi_\tau(s)$ is analytic in the
whole complex $s$-plane, except for the positive real axis, where it may have
poles and cuts. The correlator $\Pi_\tau(s)$ of hadronic QCD currents receives
contributions from vector and axialvector correlation functions for both $(ud)$
as well as $(us)$ flavour content, dressed with the appropriate quark mixing
matrix factors:
\begin{equation}
\label{PiTau}
\Pi_\tau(s) \,=\, |V_{ud}|^2\,\Big[\,\Pi_{ud}^{\rm V}(s)+\Pi_{ud}^{\rm A}(s)\,
\Big] + |V_{us}|^2\,\Big[\,\Pi_{us}^{\rm V}(s)+\Pi_{us}^{\rm A}(s)\,\Big] \,.
\end{equation}
Furthermore, $\rho_\tau(s)\equiv\IM\,\Pi_\tau(s+i0)/\pi$ is the so-called
spectral function which contains all the physical information.

In the case of hadronic $\tau$ decays, $s$ is the invariant mass of the final
hadron system, $w_\tau(s)$ is the relevant kinematic weight function,
\begin{equation}
w_\tau(s) \,=\, \frac{12\pi^2}{s_0}\biggl(1-\frac{s}{s_0}\biggr)^2
\biggl(1+2\,\frac{s}{s_0}\biggr) \,,
\end{equation}
and $s_0=M_\tau^2$, though for the moment the somewhat more general expression
will be kept. The fortuitous fact about $R_\tau$ is that the weight function
introduces a double zero at $s=s_0$, such that there is no need to evaluate
$\Pi_\tau(s)$ close to the real axis where perturbation theory becomes
problematic. In writing eq.~\eqn{RtauPi}, a small contribution from
longitudinal correlators has been omitted. Being suppressed by the light quark
masses, it is of no relevance for the purpose of this work which will
concentrate on the massless correlators, but it will be reconsidered in a
subsequent publication.

Since the hadronic $\tau$ decay rate is a physical, measurable quantity, the
corresponding perturbative QCD expression can be improved with the help of the
renormalisation group equation (RGE). In the course of calculating
$R_\tau^\Pi(s_0)$, one also has to perform the contour integration over the
circle with radius $s_0$ in the complex $s$-plane. Then the question arises in
which order both operations should be performed. First calculating the contour
integral and then performing the renormalisation group improvement goes under
the name of {\em fixed-order} perturbation theory (FOPT) whereas the second
approach of first resumming the expansion and afterwards integrating over the
contour has been suggested in \cite{piv:91,dp:92a} and is termed
{\em contour-improved} perturbation theory (CIPT). Both approaches will be
investigated in much detail below.

Numerically, it is found that the CI perturbative expansion displays a better
behaviour than the corresponding FO series \cite{dp:92a}. In particular,
employing the known perturbative results up to order $\alpha_s^3$ as well as
estimates for the contributions at order $\alpha_s^4$ and $\alpha_s^5$, it is
found that the difference between the two approaches is much larger than would
be naively expected, based on the last included terms in the expansion. In the
past, this apparent ambiguity has cast doubts as to which way of resumming the
perturbative series is preferable and it represents the dominant source of
theoretical uncertainty in the determination of $\alpha_s$ from the hadronic
$\tau$ decay rate up to this day.

Below, it is demonstrated that in fact both approaches to the renormalisation
group improvement of the perturbative series yield compatible results. However,
the fixed-order series is found to oscillate around the contour-improved result
with an oscillation frequency of approximately six perturbative orders,
approaching it until about the 30th order, after which the expansion reveals
its asymptotic nature. To this end, in the next section, the perturbative
expansion of the vector correlation function will be reviewed, and in section~3
the two approaches of improving the perturbative series in the $\tau$-decay
finite energy sum rule will be discussed. In section~4, a simplified example
will be analysed, in which only the first coefficient of the $\beta$-function
$\beta_1$ is non-vanishing. Whereas this case can be treated fully analytically,
it shares all the essential features of the complete QCD expression. The more
realistic case which includes all currently known terms in the $\beta$-function
is then treated in section~5. In section 6 and 7, the influence of variations
of the renormalisation scale as well as the renormalisation scheme on the
hadronic $\tau$ decay rate are investigated, and, finally, in the conclusions
my main findings of this work will be summarised.

\section{Perturbative correlator}

By far the most important contribution to the hadronic $\tau$ decay rate is
due to the purely perturbative vector (axialvector) correlation function
in the limit of vanishing quark masses \cite{bnp:92}. In this limit vector and
axialvector correlators yield identical contributions. Therefore, in what
follows, only the vector contribution will be investigated. A discussion of
quark mass corrections shall be presented in a forthcoming publication, but
for the remainder of this work, the quark masses are assumed to be zero.

In the massless case, the perturbative expansion of the vector correlation
function $\Pi(s)$ exhibits the general structure:\footnote{Since in
the rest of this work, we will only be concerned with the vector correlator,
the superscript V will be dropped, and also the flavour content does not matter
in the limit of vanishing quark masses.} 
\begin{equation}
\label{Pis}
\Pi(s) \,=\, -\,\frac{N_c}{12\pi^2} \sum\limits_{n=0}^\infty a_\mu^n
\sum\limits_{k=0}^{n+1} c_{nk}\,L^k  \qquad \mbox{where} \qquad L\,\equiv\,
\ln\frac{-s}{\mu^2} \,,
\end{equation}
and $a_\mu\equiv\alpha_s(\mu^2)/\pi$ with $\mu$ being the renormalisation
scale. (The global minus sign has been introduced for later convenience.)
$\Pi(s)$ itself is not a physical quantity. However, this is the case for the
spectral function $\rho(s)$ as well as for the Adler function $D(s)$:
\begin{equation}
\label{Ds}
D(s) \,\equiv\, -\,s\,\frac{d}{ds}\,\Pi(s) \,=\,
\frac{N_c}{12\pi^2} \sum\limits_{n=0}^\infty a_\mu^n
\sum\limits_{k=1}^{n+1} k\, c_{nk}\,L^{k-1} \,.
\end{equation}
It should be clear that the coefficients $c_{n0}$ do not contribute to both
the spectral function (because they are real) and to the Adler function
(because of the factor $k$). Thus the coefficients $c_{n0}$, which actually
are found to depend on the renormalisation prescription, can be considered as
``unphysical'' in that they do not appear in measurable quantities.

As physical quantities, both $D(s)$ as well as $\rho(s)$ have to satisfy a
homogeneous RGE:
\begin{equation}
\label{RGEDrho}
-\,\mu\,\frac{d}{d\mu} \Biggl\{\!\!\begin{array}{c} D(s) \\ \rho(s)
\end{array}\!\!\Biggr\} \,=\, \Biggl[\,2\,\frac{\partial}{\partial L}
+\beta(a)\,\frac{\partial}{\partial a}\,\Biggr] \Biggl\{\!\!\begin{array}{c}
D(s) \\ \rho(s) \end{array}\!\!\Biggr\} \,=\, 0 \,,
\end{equation}
where $\beta(a)$ is the QCD $\beta$-function, defined as:
\begin{equation}
\label{betaf}
-\,\mu\,\frac{da}{d\mu} \,\equiv\, \beta(a) \,=\,
\beta_1\,a^2 + \beta_2\,a^3 + \beta_3\,a^4 + \beta_4\,a^5 + \ldots \,.
\end{equation}
Numerically, for $N_c=3$ and in the $\MSb$-scheme \cite{bbdm:78} the first
four coefficients are given by \cite{tvz:80,lrv:97,cza:04}:
\begin{eqnarray}
\label{betan}
\beta_1 &=& \sfrac{11}{2} - \sfrac{1}{3}\,N_f \,, \qquad
\beta_2 \,=\, \sfrac{51}{4} - \sfrac{19}{12}\,N_f \,, \qquad
\beta_3 \,=\, \sfrac{2857}{64} - \sfrac{5033}{576}\,N_f +
              \sfrac{325}{1728}\,N_f^2
\,, \nn \\
\smvs
\beta_4 &=& \sfrac{149753}{768} + \sfrac{891}{32}\,\zeta_3 -
\Big(\sfrac{1078361}{20736} + \sfrac{1627}{864}\,\zeta_3\Big) N_f +
\Big(\sfrac{50065}{20736} + \sfrac{809}{1296}\,\zeta_3\Big) N_f^2 +
\sfrac{1093}{93312}\,N_f^3 \,. \hfill
\end{eqnarray}

The RGE puts constraints on the coefficients $c_{nk}$. Considering the
coefficients $c_{n1}$ to be independent, all other coefficients $c_{nk}$
with $k=2,\ldots,n+1$ can be expressed in terms of the $c_{n1}$ and
$\beta$-function coefficients. Up to order $\alpha_s^4$, the RG constraints
lead to:
\begin{eqnarray}
\label{cnk}
c_{22} &=& -\,\frac{\beta_1}{4}\,c_{11} \,, \quad
c_{33} \,=\, \frac{\beta_1^2}{12}\,c_{11} \,, \quad
c_{32} \,=\, -\,\frac{1}{4}\,(\beta_2\,c_{11}+2\beta_1\,c_{21}) \,, \\
\smvs
c_{44} &=& -\,\frac{\beta_1^3}{32}\,c_{11} \,, \quad
c_{43} \,=\, \frac{\beta_1}{24}\,(5\beta_2\,c_{11}+6\beta_1\,c_{21})\,, \quad
c_{42} \,=\, -\,\frac{1}{4}\,(\beta_3\,c_{11}+2\beta_2\,c_{21}+
3\beta_1\,c_{31}) \,. \nn
\end{eqnarray}
Furthermore, the coefficients $c_{n,n+1}=0$ for $n\geq 1$. The independent
coefficients $c_{n1}$ are known analytically up to order $\alpha_s^3$
\cite{gkl:91,ss:91} and at $N_c=3$ in the $\MSb$-scheme take the following
values:
\begin{eqnarray}
\label{cn1}
c_{01} &=& c_{11} \,=\, 1 \,, \quad 
c_{21} \,=\, \sfrac{365}{24}-11\zeta_3-
\Big(\sfrac{11}{12}-\sfrac{2}{3}\zeta_3\Big) N_f \,, \nn \\
\smvs
c_{31} &=& \sfrac{87029}{288}-\sfrac{1103}{4}\zeta_3+\sfrac{275}{6}
\zeta_5 - \Big(\sfrac{7847}{216}-\sfrac{262}{9}\zeta_3+\sfrac{25}{9}\zeta_5
\Big) N_f + \Big(\sfrac{151}{162}-\sfrac{19}{27}\zeta_3\Big) N_f^2 \,.
\end{eqnarray}
For the next five- and six-loop coefficients $c_{41}$ and $c_{51}$, estimates
employing principles of ``minimal sensitivity'' (PMS) or ``fastest apparent
convergence'' (FAC) \cite{ste:81,pen:82}, together with recently computed
terms of order $\alpha_s^4\,N_f^2$, exist which read \cite{ks:95,bck:02}:
\begin{equation}
\label{c41c51}
c_{41} \,=\,  27 \pm  16 \,, \qquad 
c_{51} \,=\, 145 \pm 100 \,.
\end{equation}
For illustrative purposes, also the central values of these estimates will
be taken into account in the analysis below.

Since the Adler function $D(s)$ satisfies a homogeneous RGE, the logarithms
in eq.~\eqn{Ds} can be resummed with the choice $\mu^2=-s\equiv Q^2$, leading
to the simple expression:
\begin{equation}
\label{Dsresum}
D(Q^2) \,=\, \frac{N_c}{12\pi^2} \sum\limits_{n=0}^\infty c_{n1}\,a_Q^n \,,
\end{equation}
where $a_Q\equiv\alpha_s(Q^2)/\pi$. As is again apparent from this equation,
the only physically relevant coefficients are the $c_{n1}$.

\section{Hadronic \boldmath{$\tau$} decay rate}

In principle, now one could proceed by inserting the general expression
\eqn{Pis} for $\Pi(s)$ into the contour integral of eq.~\eqn{RtauPi}. However,
it is advantageous to rewrite the FESR for the $\tau$ decay rate in terms of
the Adler function $D(s)$ by partial integration:\footnote{In writing
eq.~\eqn{RtauD}, $|V_{ud}|^2+|V_{us}|^2=1$ has been assumed, which numerically
is satisfied rather well.}
\begin{equation}
\label{RtauD}
R_\tau^D(s_0) \,=\, -\,6\pi i \!\!\oint\limits_{|x|=1} \!\!\frac{dx}{x}\,
(1-x)^3\,(1+x)\,D(s_0 x) \,,
\end{equation}
with the new dimensionless integration variable being $x\equiv s/s_0$. For
FOPT, order by order $R_\tau^\Pi(s_0)$ and $R_\tau^D(s_0)$ can be shown to be
identical as they should, but applying CIPT, the perturbative expansion for
$R_\tau^\Pi(s_0)$ is less well behaved than the one for $R_\tau^D(s_0)$.
Therefore, only the comparison of $R_\tau^D(s_0)$ in FOPT and CIPT will be
investigated in this work.

Inserting the expression \eqn{Ds} for $D(s)$ into the contour integral of
eq.~\eqn{RtauD}, one finds:
\begin{eqnarray}
\label{RtauD1}
R_\tau^D(s_0) &=& \frac{N_c}{2\pi i}\, \sum\limits_{n=0}^\infty
a_\mu^n \sum\limits_{k=1}^{n+1} k\,c_{nk} \!\!\oint\limits_{|x|=1} \!\!
\frac{dx}{x}\, (1-x)^3\,(1+x) \ln^{k-1}\biggl(\frac{-s_0 x}{\mu^2}\biggr) \nn\\
\smvs
&=& \frac{N_c}{2\pi i}\, \sum\limits_{n=0}^\infty a_\mu^n
\sum\limits_{k=1}^{n+1} k\,c_{nk} \sum\limits_{l=0}^{k-1} {k-1 \choose l}
\ln^{k-l-1}\frac{s_0}{\mu^2}\!\! \oint\limits_{|x|=1} \!\! \frac{dx}{x}\,
(1-x)^3\,(1+x) \ln^l(-x) \nn \\
\smvs
&=& N_c \sum\limits_{n=0}^\infty a_\mu^n\sum\limits_{k=1}^{n+1}
k\,c_{nk} \sum\limits_{l=0}^{k-1} {k-1 \choose l} J_l\,\ln^{k-l-1}\frac{s_0}
{\mu^2} \,.
\end{eqnarray}
In the last line, the contour integrals $J_l$ are defined by
\begin{equation}
\label{Jl}
J_l \,\equiv\, \frac{1}{2\pi i} \!\!\oint\limits_{|x|=1} \!\!
\frac{dx}{x}\, (1-x)^3\,(1+x) \ln^l(-x) \,=\,
\frac{1}{2\pi}\,\Big[\, I_{l,0} + 2\,I_{l,1} - 2\,I_{l,3} - I_{l,4} \,\Big] \,,
\end{equation}
with the required integrals $I_{l,m}$ being given by
\begin{eqnarray}
\label{Ilm}
I_{l,m} \,\equiv\,
i^l \!\!\int\limits_{-\pi}^{+\pi} \alpha^l\,{\rm e}^{im\alpha}\,d\alpha &=&
i\,\biggl(\frac{-1}{m}\biggr)^{\!l+1}\Gamma(l+1,-i\alpha m)\biggr|_{-\pi}
^{+\pi} \nn \\
\smvs
&=& (-1)^{l+m}\frac{2\,l!}{m^{l+2}}\!\sum\limits_{k=1}^{[(l+1)/2]}
(-1)^k \,\frac{m^{2k} \pi^{2k-1}}{(2k-1)!} \,,
\end{eqnarray}
where $\Gamma(l+1,z)$ is the incomplete $\Gamma$-function, $[n]$ denotes the
integer part of $n$ and $m\geq1$. For the case $m=0$, one obtains
$I_{l,0}\,=\,i^l[1+(-1)^l]\,\pi^{l+1}/(l+1)$. Explicitly, the first few of the
integrals $J_l$, which are needed up to order $\alpha_s^3$, read:
\begin{equation}
\label{J0to3}
J_0 \,=\, 1 \,, \quad
J_1 \,=\, -\,\sfrac{19}{12} \,, \quad
J_2 \,=\, \sfrac{265}{72} - \sfrac{1}{3}\,\pi^2 \,, \quad
J_3 \,=\, -\,\sfrac{3355}{288} + \sfrac{19}{12}\,\pi^2 \,,
\end{equation}
in agreement with ref.~\cite{dp:92a}.

As discussed above, the Adler function $D(s)$ is a physical quantity, and thus
satisfies a homogeneous RGE. Therefore, the logarithms in eq.~\eqn{RtauD1}
can be resummed. Let us first concentrate on FO perturbation theory, which
amounts to taking the choice $\mu^2=s_0$ in the last line of \eqn{RtauD1},
which then leads to
\begin{equation}
\label{RtauDFO}
R_\tau^{D,{\rm FO}}(s_0) \,=\, N_c \sum\limits_{n=0}^\infty a_{s_0}^n
\sum\limits_{k=1}^{n+1} k\,c_{nk}\,J_{k-1} \,,
\end{equation}
with $a_{s_0}\equiv a(s_0)$. Inserting the relations \eqn{cnk} for the $c_{nk}$
together with the corresponding contour integrals $J_{k-1}$, up to order
$a_{s_0}^4$, this leads to:
\begin{eqnarray}
\label{RtauDFO1}
R_\tau^{D,{\rm FO}}(s_0) &=& N_c\,\Biggl\{\, c_{01} + c_{11}\,a_{s_0}
+ \Big( c_{21} + \sfrac{19}{24}\beta_1 c_{11}\Big) a_{s_0}^2 \nn \\
\smvs
&& + \biggl( c_{31} + \sfrac{19}{12}\beta_1 c_{21} + \Big( \sfrac{19}{24}
\beta_2 + \Big( \sfrac{265}{288} - \sfrac{\pi^2}{12} \Big)\beta_1^2 \Big)
c_{11} \biggr) a_{s_0}^3 \nn \\
\smvs
&& +\,\biggl( c_{41} + \sfrac{19}{8}\beta_1 c_{31} + \Big( \sfrac{19}{12}
\beta_2 + \Big( \sfrac{265}{96}-\sfrac{\pi^2}{4} \Big)\beta_1^2 \Big) c_{21}
\nn \\
\smvs
&& \hspace{12.4mm} +\,\Big( \sfrac{19}{24}\beta_3 + \Big( \sfrac{1325}{576} -
\sfrac{5\pi^2}{24} \Big)\beta_2\beta_1 + \Big( \sfrac{3355}{2304} -
\sfrac{19}{96}\pi^2 \Big) \beta_1^3 \Big) c_{11} \biggr) a_{s_0}^4 \,\Biggr\}\\
\smvs
\label{RtauDFO2}
&=& N_c\,\Biggl\{\, 1 + a_{s_0} + \Big(\sfrac{769}{48}-9\,
\zeta_3\Big) a_{s_0}^2 + \Big(\sfrac{363247}{1152}-\sfrac{27}{16}\,\pi^2-
\sfrac{2071}{8}\,\zeta_3+\sfrac{75}{2}\,\zeta_5\Big) a_{s_0}^3 \nn \\
\smvs
&& +\,\Big(\sfrac{19907171}{6144}+c_{41}-\sfrac{22683}{256}\,\pi^2-\Big(
\sfrac{345405}{128}-\sfrac{729}{16}\,\pi^2\Big)\zeta_3+\sfrac{12825}{32}\,
\zeta_5\Big) a_{s_0}^4 \,\Biggr\} \,.
\end{eqnarray}
For the second step, the coefficients of the $\beta$-function \eqn{betan} as
well as the $c_{n1}$-coefficients of eq.~\eqn{cn1} have been employed. An
important remark is in order here. As should be clear from eq.~\eqn{RtauDFO1},
each order $n$ in FOPT depends on all coefficients $c_{i1}$ with $i\leq n$.
This will play a crucial role in what follows. Before analysing the expressions
numerically, the second approach of resumming the logarithms, namely CI
perturbation theory, will be introduced.

For CIPT, the logarithms in eq.~\eqn{RtauD1} should be resummed before
performing the contour integral. This can be achieved with the choice
$\mu^2=-s_0 x$ in the first line of \eqn{RtauD1}, which yields
\begin{equation}
\label{RtauDCI}
R_\tau^{D,{\rm CI}}(s_0) \,=\, N_c \sum\limits_{n=0}^\infty
c_{n1}\,J_n^a(s_0) \,.
\end{equation}
Here, the equation has been rewritten in terms of the contour integrals
$J_n^a(s_0)$, defined as:
\begin{equation}
\label{Jna}
J_n^a(s_0) \,\equiv\, \frac{1}{2\pi i} \!\!\oint\limits_{|x|=1}\!\!
\frac{dx}{x}\,(1-x)^3\,(1+x)\,a^n(-s_0 x) \,.
\end{equation}
In contrast to the FO case, for CIPT each order $n$ just depends on the
corresponding coefficient $c_{n1}$. Thus, all contributions proportional to
the coefficient $c_{n1}$ which in FOPT appear at all perturbative orders
equal or greater to $n$ are resummed into a single term. This is related to the
fact that CIPT resums the running of the QCD coupling along the integration
contour in the complex $s$-plane. To perform the contour integration, we have
to analytically continue the strong coupling $\alpha_s$ to the complex
$s$-plane, but this is straightforward since the dependence of $\alpha_s$ on
$s$ is only logarithmic and it has the same cut structure as $\Pi(s)$.

\section{A simple example}

Before embarking on the general integrals $J_n^a(s_0)$, as a first step, let
us investigate a simplified example which can be treated fully analytically,
namely the case where only the first coefficient of the $\beta$-function,
$\beta_1$, is non-zero, and all higher order $\beta$-coefficients vanish
identically. Then the running coupling in eq.~\eqn{Jna} can be expressed as
\begin{equation}
\label{as0xb1}
a^n(-s_0 x) \,=\, \frac{a_{s_0}^n}{\Big[1+\sfrac{\beta_1}{2}\,a_{s_0}
\ln(-x)\Big]^n} \,,
\end{equation}
and the integrals $J_n^a(s_0)$ can be calculated analytically with the result:
\begin{equation}
\label{Jnaa}
J_n^a(s_0) \,=\, a_{s_0}^n\,\Big[\, J_{n,0}(\kappa) + 2 J_{n,1}(\kappa) -
2 J_{n,3}(\kappa) - J_{n,4}(\kappa)\,\Big] \,.
\end{equation}
Here, the constant $\kappa\equiv \beta_1 a_{s_0}/2$, and the integrals
$J_{n,m}(\kappa)$ are found to be
\begin{equation}
\label{Jnm}
J_{n,m}(\kappa) \,=\, \frac{{\rm e}^{-\frac{m}{\kappa}} \Big(
\sfrac{m}{\kappa}\Big)^{n-1}}{2\pi i\kappa\,\Gamma(n)}\Biggl[\,{\rm Ei}\Big(
\sfrac{m}{\kappa} z\Big) - {\rm e}^{\frac{m}{\kappa}z} \sum\limits_{l=1}^{n-1}
\Gamma(l)\Big(\sfrac{m}{\kappa} z\Big)^{-l}\,
\Biggr]_{z=1-i\pi\kappa}^{z=1+i\pi\kappa} \,,
\end{equation}
with ${\rm Ei}(z)=\int {\rm e}^z/z\,dz$ being the exponential-integral
function and $m\geq 1$. Again, the case $m=0$ which is also required has to
be treated separately and yields:
\begin{eqnarray}
\label{J10}
J_{1,0}(\kappa) &=& \frac{1}{2\pi i\kappa}\,
\ln\frac{(1+i\pi\kappa)}{(1-i\pi\kappa)} \,,\quad n=1 \,, \\
\smvs
\label{Jn0}
J_{n,0}(\kappa) &=& \frac{1}{2\pi i\kappa(n-1)}\,
\Big[\,(1-i\pi\kappa)^{1-n} - (1+i\pi\kappa)^{1-n} \,\Big] \,,\quad n\geq 2 \,.
\end{eqnarray}

The analytic expressions for the integrals $J_n^a(s_0)$ can readily be compared
with their perturbative expansion which is obtained by expanding
eq.~\eqn{as0xb1} in powers of $\kappa=\beta_1 a_{s_0}/2$, before performing the
contour integration:
\begin{equation}
\label{Jnae}
J_n^a(s_0) \,=\, \frac{a_{s_0}^n}{\Gamma(n)}\,
\sum\limits_{l=0}^{\infty} \,\frac{\Gamma(n+l)}{l!}\,J_l\,(-\kappa)^l \,.
\end{equation}
Employing the asymptotic expansion
for the exponential-integral function in eq.~\eqn{Jnm} as well as expanding
eqs.~\eqn{Jnm} to \eqn{Jn0} in powers of $\kappa$, one can verify the agreement
between both representations for $J_n^a(s_0)$.

Let us now turn to a numerical comparison of CI versus FO perturbation theory
in the simplified example. To do this, an input value for $\alpha_s$ is
required and the very recent result $\alpha_s(M_\tau^2)= 0.34$ \cite{aleph:05}
will be employed for $a_{s_0}$. The CI expression of eq.~\eqn{Jnaa} and the
perturbative expansion of the FO result of eq.~\eqn{Jnae}, together with
eq.~\eqn{RtauDCI}, then lead to:
\begin{eqnarray}
&& \hspace{5mm} a^0 \hspace{8mm} a^1 \hspace{11.6mm} a^2 \hspace{11.6mm} a^3
\hspace{11.6mm} a^4 \hspace{11.6mm} a^5 \nn \\
\label{RtauCIb1n}
R_{\tau,\,\beta_1}^{D,{\rm CI}} &=& 3\,\Big[\,
1 + 0.1455 + 0.0305 + 0.0136 + 0.0060 + 0.0031 \,\Big] \,=\, 3.5961 \,, \\
\smvs
\label{RtauFOb1n}
R_{\tau,\,\beta_1}^{D,{\rm FO}} &=& 3\,\Big[\,
1 + 0.1082 + 0.0609 + 0.0254 + 0.0082 + 0.0024 \,\Big] \,=\, 3.6154 \,.
\end{eqnarray}
From \eqn{RtauCIb1n} and \eqn{RtauFOb1n} one observes that the CI expansion
appears to display a better convergence, although the last included term of
the FO expansion is somewhat smaller. Altogether, the difference between the
two results amounts to $3\cdot 0.0064$, about 2-3 times the last term in both
expansions. Thus, the agreement between both approaches to performing the
renormalisation group improvement is reasonably good.

\TABLE[ht]{%
\begin{tabular}{rrrrrrrr}
\hline\hline
 1\po &  2\po &  3\po &  4\po &  5\po &  6\po &  7\po &  8\po \\
\s{ 0.108225} & \s{ 0.060933} & \s{ 0.025394} & \s{ 0.008165} &
\s{ 0.002419} & \s{-0.000607} & \s{-0.004193} & \s{-0.004669} \\
\hline
 9\po & 10\po & 11\po & 12\po & 13\po & 14\po & 15\po & 16\po \\
\s{-0.000104} & \s{ 0.003728} & \s{ 0.001815} & \s{-0.001991} &
\s{-0.001863} & \s{ 0.000850} & \s{ 0.001407} & \s{-0.000275} \\
\hline
17\po & 18\po & 19\po & 20\po & 21\po & 22\po & 23\po & 24\po \\
\s{-0.000939} & \s{ 0.000032} & \s{ 0.000588} & \s{ 0.000051} &
\s{-0.000354} & \s{-0.000066} & \s{ 0.000209} & \s{ 0.000057} \\
\hline
25\po & 26\po & 27\po & 28\po & 29\po & 30\po & 31\po & 32\po \\
\s{-0.000122} & \s{-0.000043} & \s{ 0.000070} & \s{ 0.000030} &
\s{-0.000040} & \s{-0.000020} & \s{ 0.000023} & \s{ 0.000013} \\
\hline\hline
\end{tabular}
\caption{Perturbative order $n$ and corresponding terms in the expansion of
 $R_{\tau,\,\beta_1}^{D,{\rm FO}}/3$.
\label{tab1}}}

Nevertheless, if we were to assume that all coefficients $c_{n1}$ with
$n\geq 6$ vanish identically, the CI result of eq.~\eqn{RtauCIb1n} would
represent the complete answer. Then the question arises, how the FO result
approaches this value at higher orders in perturbation theory. Since with
eq.~\eqn{Jnae} an all order expression is at our disposal, the question can
be answered by inserting \eqn{Jnae} into eq.~\eqn{RtauDCI} and reexpanding in
powers of $a_{s_0}$. The numerical result of this exercise is displayed in
table~\ref{tab1} up to the 32nd order.  The entries are the perturbative order
$n$ at the one hand and the corresponding term in the expansion of
$R_{\tau,\,\beta_1}^{D,{\rm FO}}/3$ on the other. Therefore, the first five
entries agree with the corresponding terms in the square bracket of
eq.~\eqn{RtauFOb1n}.

\FIGURE[hb]{\includegraphics[angle=0, width=14cm]{diffFOb1}
\caption{Difference
$(R_{\tau,\,\beta_1}^{D,{\rm FO}}-R_{\tau,\,\beta_1}^{D,{\rm CI}})/3$ as a
function of the perturbative order $n$ up to which the terms in the FOPT
series are summed in the example with only $\beta_1$ non-vanishing.
\label{fig1}}}

By inspecting table~\ref{tab1}, it is observed that after the first five
positive terms, there are four terms with negative sign. After that, the
sign changes at about every second term. Generally, the size of the terms
decreases with increasing order, but not from one term to the next. A graphical
account of these findings is presented in figure~\ref{fig1}, which shows the
difference $(R_{\tau,\,\beta_1}^{D,{\rm FO}}-R_{\tau,\,\beta_1}^{D,{\rm CI}})/3$
as a function of the perturbative order $n$ up to which the terms in the FOPT
series are summed. As can be seen from this figure, the sign changes lead to
FO results which oscillate around the CI value. Due to the sign change at about
every second term, the oscillation frequency is approximately four perturbative
orders. Furthermore, the magnitude of the oscillation is tending towards zero,
such that the FO value approaches the CI result at large orders. Incidentally,
the maximum of the difference between both approaches is found to appear at the
fifth order. Calculating the series up to the 200th order, the series still
happens to be convergent, and no sign of a divergent behaviour is observed.
As has already been discussed in \cite{dp:92a}, for $\kappa<1/\pi$, which
corresponds to $\alpha_s(M_\tau^2)<2/\beta_1=0.44$, the expansion \eqn{as0xb1}
converges on the whole unit circle. Thus, also the FO series for
$R_{\tau,\,\beta_1}^{D,{\rm FO}}$ should be convergent for the physical value
of $\alpha_s(M_\tau^2)$, at least if only a finite number of $c_{n1}$
coefficients are included.

\section{The general case}

In order to tackle the general problem with all $\beta$-function coefficients
$\beta_1$ to $\beta_4$ being unequal to zero, the general expansion of
$a(-s_0 x)$ in terms of $a_{s_0}$ is employed, which takes the form:
\begin{equation}
\label{as0x}
a(-s_0 x) \,=\, a_{s_0}\,\biggl[\, 1 + \sum\limits_{i=1}^\infty \,
\sum\limits_{j=1}^i \, d_{ij}\,a_{s_0}^i \ln^j(-x) \,\biggr] \,.
\end{equation}
The coefficients $d_{ij}$ can be calculated from the RGE for the QCD coupling
eq.~\eqn{betaf}. The coefficients of the highest and the lowest power of the
logarithm can easily be given for arbitrary order and read
$d_{nn}=(-\beta_1/2)^n$ and $d_{n1}=-\beta_n/2$. The remaining three
coefficients up to fourth order are found to be:
\begin{equation}
\label{dij}
d_{32} \,=\, \sfrac{5}{8}\,\beta_1\,\beta_2 \,, \qquad
d_{43} \,=\, -\,\sfrac{13}{24}\,\beta_1^2\,\beta_2 \,, \qquad
d_{42} \,=\, \sfrac{3}{8}\,(\beta_2^2+2\beta_1\,\beta_3) \,.
\end{equation}
In principle -- given the necessary computing resources -- the coefficients
$d_{ij}$ can be calculated to an arbitrary order, taking into account the
known $\beta$-function coefficients, and setting the even higher-order ones
to zero.

The numerical analysis of the case including all available $\beta$-function
coefficients proceeds along similar lines as the simplified example in the
previous section. For CIPT, now the integrals of eq.~\eqn{Jna} have to be
calculated numerically by plugging a numerical solution of the RGE \eqn{betaf}
for the running coupling and then performing the contour integration. The FO
result can be taken from eq.~\eqn{RtauDFO1}. In addition including the
fifth-order term, one obtains:
\begin{eqnarray}
&& \hspace{5mm} a^0 \hspace{8mm} a^1 \hspace{11.6mm} a^2 \hspace{11.6mm} a^3
\hspace{11.6mm} a^4 \hspace{11.6mm} a^5 \nn \\
\label{RtauCIn}
R_{\tau}^{D,{\rm CI}} &=& 3\,\Big[\,
1 + 0.1479 + 0.0297 + 0.0122 + 0.0047 + 0.0019 \,\Big] \,=\, 3.5893 \,, \\
\smvs
\label{RtauFOn}
R_{\tau}^{D,{\rm FO}} &=& 3\,\Big[\,
1 + 0.1082 + 0.0609 + 0.0334 + 0.0144 + 0.0021 \,\Big] \,=\, 3.6571 \,.
\end{eqnarray}
Again, the CI series displays a better convergence although the last included
term in both expansions is almost of the same size. This time, however, the
difference between CI and FO perturbation theory looks much more dramatic, as
it amounts to $3\cdot 0.0226$, more than ten times the last included term! This
apparent ambiguity in the perturbative prediction of the hadronic $\tau$ decay
rate at the moment represents the dominant theoretical uncertainty in the
extraction of the strong coupling $\alpha_s$ from this channel.

\TABLE[ht]{\begin{tabular}{rrrrrrrr}
\hline\hline
 1\po &  2\po &  3\po &  4\po &  5\po &  6\po &  7\po &  8\po \\
\s{ 0.108225} & \s{ 0.060933} & \s{ 0.033422} & \s{ 0.014405} &
\s{ 0.002052} & \s{-0.006847} & \s{-0.012518} & \s{-0.012029} \\
\hline
 9\po & 10\po & 11\po & 12\po & 13\po & 14\po & 15\po & 16\po \\
\s{-0.004380} & \s{ 0.005429} & \s{ 0.010012} & \s{ 0.006522} &
\s{-0.001223} & \s{-0.006793} & \s{-0.006409} & \s{-0.001148} \\
\hline
17\po & 18\po & 19\po & 20\po & 21\po & 22\po & 23\po & 24\po \\
\s{ 0.004472} & \s{ 0.005997} & \s{ 0.002420} & \s{-0.003022} &
\s{-0.005577} & \s{-0.003107} & \s{ 0.002045} & \s{ 0.005211} \\
\hline
25\po & 26\po & 27\po & 28\po & 29\po & 30\po & 31\po & 32\po \\
\s{ 0.003582} & \s{-0.001304} & \s{-0.004970} & \s{-0.004026} &
\s{ 0.000713} & \s{ 0.004875} & \s{ 0.004498} & \s{-0.000213} \\
\hline\hline
\end{tabular}
\caption{Perturbative order $n$ and corresponding terms in the expansion of
 $R_{\tau}^{D,{\rm FO}}/3$.
\label{tab2}}}

\FIGURE[b]{\includegraphics[angle=0, width=14cm]{diffFO}
\caption{Difference
$(R_{\tau}^{D,{\rm FO}}-R_{\tau}^{D,{\rm CI}})/3$ as a function of the
perturbative order $n$ up to which the terms in the FOPT series are summed.
\label{fig2}}}

To gain further insight into the origin of the problem, like in the previous,
simplified example, higher order contributions should be inspected. These can
be acquired by exploiting the expansion \eqn{as0x} for the running coupling in
the contour integrals of eq.~\eqn{Jna} which can be expressed in terms of the
$J_l$, and then reexpanding the result in terms of $a_{s_0}$.\footnote{Of
course, this is just another way of calculating higher order relations for the
$c_{nk}$ coefficients, similar to the ones given in eq.~\eqn{cnk}.} The result
of this exercise is presented in table~\ref{tab2}. Like in the last section,
after the first five positive terms, there are four negative ones after which
the terms change sign at every third order, and, all in all, decrease in
magnitude. The negative sign of the sixth order term, which to some extent
already lowers the difference between CIPT and FOPT, has also been observed
recently in ref.~\cite{dhz:05}.

Restricting oneself to the five $c_{n1}$ coefficients discussed above as well
as the four known $\beta$--function coefficients, again the CI value
\eqn{RtauCIn} represents the full answer. This time, in figure~\ref{fig2} the
difference between the FO expansion summed up to a particular order $n$ and
the complete CI result is displayed up to the 51st order. Until about the 30th
order, the series shows the same behaviour as in the simplified example above.
It oscillates around the CI result, while approaching it. However, for the
general case the convergence to the CI result is much slower, and roughly after
the 30th order, an asymptotic behaviour of the series sets in and the terms
start to again increase. The observed asymptotic nature of the FOPT series for
$R_{\tau}^{D,{\rm FO}}$ has nothing to do with the expected asymptotic
behaviour of the $c_{n1}$ coefficients which is due to renormalons
\cite{bbb:95,ben:98}. Inspection of the expansion \eqn{as0x} reveals that for
the used physical value of $\alpha_s(M_\tau^2)$, it diverges on part of the
unit circle, leading to the behaviour detected in figure~\ref{fig2}. The
minimal deviation of the FO result at the 30th order is about $0.005$, and the
oscillation frequency is approximately six perturbative orders. Alas, again
the maximum of the deviation occurs at the fifth order.

\section{Scale variations}

In perturbative QCD calculations, it is quite standard to estimate uncertainties
due to as yet uncomputed higher orders by a variation of the in principle
arbitrary renormalisation scale. Thus, in this section the influence of scale
variations on FO and CI perturbation theory for the hadronic $\tau$ decay rate
shall be investigated. Let us begin with a study of renormalisation group
improved FOPT.

Like in ref.~\cite{dp:92a}, the arbitrary renormalisation scale will be
introduced by substituting $\mu^2=\xi^2 s_0$ in the last line of
eq.~\eqn{RtauD1}, which yields:
\begin{equation}
\label{RtauDFOksi}
R_\tau^{D,{\rm FO}}(\xi^2 s_0) \,=\,
N_c \sum\limits_{n=0}^\infty a^n(\xi^2 s_0)\sum\limits_{k=1}^{n+1}
k\,c_{nk} \sum\limits_{l=0}^{k-1} {k-1 \choose l} J_l\,(-2\ln\xi)^{k-l-1}
 \,.
\end{equation}
The previous eq.~\eqn{RtauDFO} just corresponds to the special case $\xi=1$.
One can then proceed with a numerical analysis, completely analogous to the
presentation in the last section. The result of this exercise is shown in
figure~\ref{fig3} up to the 60th perturbative order and for two values of
$\xi$. The full triangles correspond to the case $\xi=0.9$ whereas the full
circles result by setting $\xi=1.1$. To guide the eye, the data points have
been connected by straight line segments. The required inputs for
$a(\xi^2 M_\tau^2)$ have been calculated by solving the RGE \eqn{betaf} with
the initial value $\alpha_s(M_\tau^2)= 0.34$.

\FIGURE[t]{\includegraphics[angle=0, width=14cm]{diffFOksi}
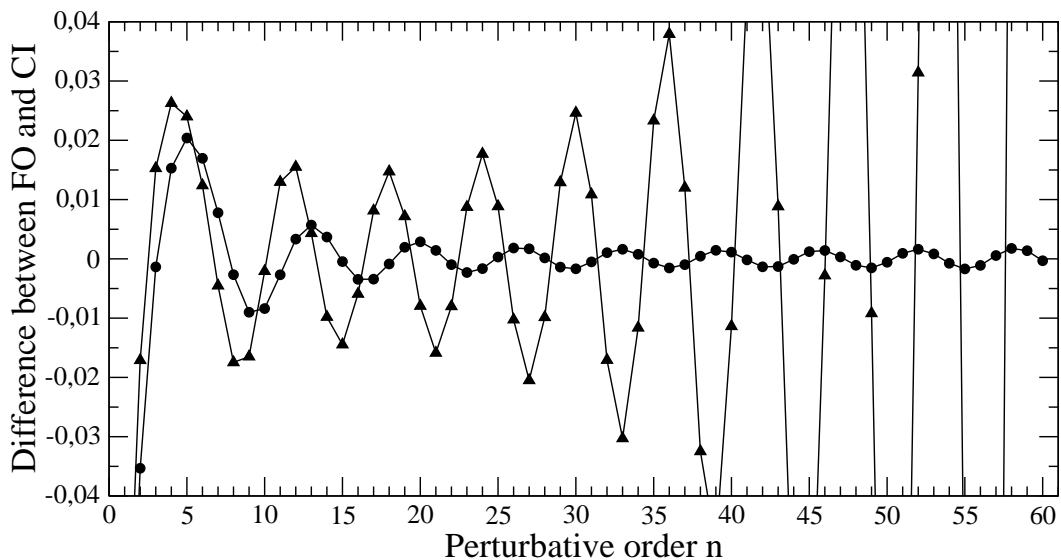
\caption{Difference
$(R_{\tau}^{D,{\rm FO}}-R_{\tau}^{D,{\rm CI}})/3$ as a function of the
perturbative order $n$ up to which the terms in the FOPT series are summed
and for an arbitrary renormalisation scale parameter $\xi$. Full triangles
correspond to $\xi=0.9$; full circles to $\xi=1.1$.
\label{fig3}}}

As can be observed from figure~\ref{fig3}, changing the renormalisation scale
even by only a small amount, the behaviour of the FO series changes quite
drastically. In the case of a smaller renormalisation scale $\xi=0.9$, the
asymptotic behaviour already sets in around the 15th order, and the minimal
deviation from the CI result is as large as $0.014$, while for the case of
a larger renormalisation scale $\xi=1.1$, close inspection shows that the
amplitude of the oscillations decreases until about the 45th order after
which it gradually again increases. This time the minimal amplitude turns
out to be $0.0013$. Like in the case $\xi=1$, however, for both examples the
oscillation frequency is about six perturbative orders. Therefore, again for
values of the renormalisation scale different from $\xi=1$, the FOPT series is
found to be compatible with the CIPT value \eqn{RtauCIn}, although due to its
rather bad behaviour it should not be utilised for any phenomenological
analysis.

For CIPT, introducing an arbitrary renormalisation scale parameter $\xi$
can be achieved with the choice $\mu^2=-\xi^2 s_0 x$ in the first line of
\eqn{RtauD1}, which results in
\begin{equation}
\label{RtauDCIksi}
R_\tau^{D,{\rm CI}}(\xi^2 s_0) \,=\, N_c \sum\limits_{n=0}^\infty
J_n^a(\xi^2 s_0) \sum\limits_{k=1}^{n+1} k\,c_{nk} (-2\ln\xi)^{k-1} \,.
\end{equation}
This time, the equation has been rewritten in terms of the contour integrals
$J_n^a(\xi^2 s_0)$, given by:
\begin{equation}
\label{Jnaksi}
J_n^a(\xi^2 s_0) \,\equiv\, \frac{1}{2\pi i} \!\!\oint\limits_{|x|=1}\!\!
\frac{dx}{x}\,(1-x)^3\,(1+x)\,a^n(-\xi^2 s_0 x) \,.
\end{equation}
From eq.~\eqn{RtauDCIksi} it is apparent that in contrast to CIPT with $\xi=1$
of eq.~\eqn{RtauDCI}, and similarly to FOPT, at a given order $n$, via the
coefficients $c_{nk}$, all independent coefficients $c_{i1}$ with $i\leq n$
contribute. Even if again only the first five $c_{n1}$ are assumed to be
non-zero, nevertheless, the CIPT series for general $\xi$ does not terminate
at this order. Hence, now also for CIPT one can investigate the difference
between $R_\tau^{D,{\rm CI}}(\xi^2 s_0)$ for a given $\xi$ and the complete
result $R_\tau^{D,{\rm CI}}$ of eq.~\eqn{RtauCIn}.

\FIGURE[t]{\includegraphics[angle=0, width=14cm]{diffCIksi}
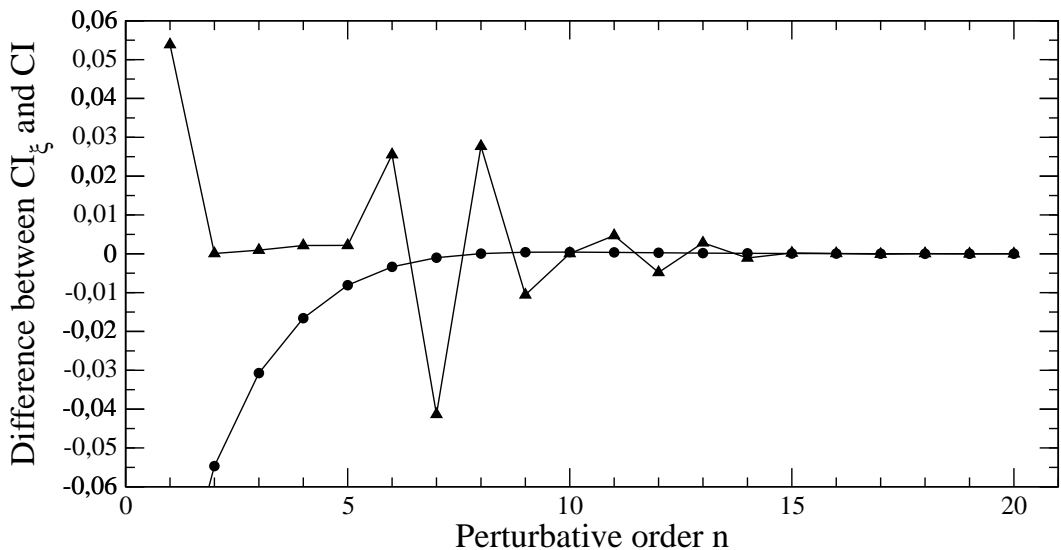
\caption{Difference
$(R_{\tau}^{D,{\rm CI}}(\xi^2 s_0)-R_{\tau}^{D,{\rm CI}})/3$ as a function of
the perturbative order $n$ up to which the terms in the CIPT series are summed
and for an arbitrary renormalisation scale parameter $\xi$. Full triangles
correspond to $\xi=0.5$; full circles to $\xi=2.0$.
\label{fig4}}}

The result of this exercise is displayed in figure~\ref{fig4} up to the 20th
perturbative order. The full triangles now correspond to the case $\xi=0.5$
whereas the full circles result by setting $\xi=2.0$. To guide the eye, the
data points have been connected by straight line segments. Generally, the
CIPT for arbitrary $\xi$ is much more stable against a variation of the scale
parameter than the corresponding FO series. Thus, a larger variation of $\xi$
could be considered. In the case of the smaller $\xi=0.5$, surprisingly, and
most probably by chance, between the second and the fifth order,
eq.~\eqn{RtauDCIksi} already gives a rather good approximation to the full
answer. Beginning with the sixth order, the deviation again fluctuates more
strongly until it finally converges towards the result of eq.~\eqn{RtauCIn}
for even higher orders. In the case of a larger scale with $\xi=2.0$, the
approach towards the complete result is much more smooth. Coming from below,
the difference only assumes small positive values before again approaching
zero. Nevertheless, like for utilising FOPT, there is no real incentive to
employ CIPT with values for $\xi$ different from one. This procedure only
reshuffles the perturbative expansion, thereby transferring known contributions
to higher orders which can as well be resummed.

After the discussion of renormalisation scale dependence of FOPT and CIPT,
let us now address the important question which are the uncertainties of the
result \eqn{RtauCIn} and how they might be estimated. Obviously, one source
of uncertainty resides in the contribution of yet uncalculated higher order
coefficients $c_{n1}$. Schemes to estimate these coefficients have been
discussed in the literature \cite{ste:81,pen:82,ks:95,bck:02} (and references
therein),\footnote{Such estimates for the coefficients $c_{41}$ and $c_{51}$
had been employed above.} but it seems fair to say that no rigorous assessment
of the corresponding uncertainty exists, prior to an explicit calculation of a
certain coefficient. It also remains unclear if using the size of the last
computed term as an estimation for yet higher order contributions is
justified -- although this is often done in practice. On the other hand, a
variation of the renormalisation scale only amounts to a reordering of the
perturbative expansion for a given set of known $c_{n1}$ coefficients and has
nothing to do with the uncertainty due to so far uncalculated coefficients.
This is somewhat different for a change of the renormalisation scheme, as will
be discussed in the next section.

A second source of uncertainty in CIPT results from the fact that the
running of the QCD coupling is required while integrating over the contour
in the complex $s$-plane, and the perturbative expansion for the
$\beta$-function has been truncated at the fourth order. An indication of
the corresponding uncertainty can be gained by comparing the result
\eqn{RtauCIn} with the value that would be obtained while using three-loop
running for $\alpha_s$. Numerically, the difference between both approaches
is found to be $3\cdot 0.00099$, only about half of the last included term
in the perturbative series. The third, and final uncertainty on the result
\eqn{RtauCIn} comes from the error of the input value for $\alpha_s(M_\tau^2)$.
This uncertainty, however, can be worked out straightforwardly, assuming a
given error on $\alpha_s(M_\tau^2)$.

\section{Scheme variations}

Besides investigating the dependence of the perturbative expansion on the
choice of the renormalisation scale, also a modification of the renormalisation
scheme provides interesting further information. In analogy to the general
expression \eqn{as0x} for the shift of the scale at which the QCD coupling is
evaluated, a transformation of the renormalisation scheme can be represented
by the following equation:
\begin{equation}
\label{ats0}
\tilde a_{s_0} \,=\, a_{s_0}\,\biggl[\, 1 + \sum\limits_{i=1}^\infty \,
d_{i}\,a_{s_0}^i \,\biggr] \,.
\end{equation}
As before, initial quantities, like the coupling $a_{s_0}$, are given in the
$\MSb$-scheme, whereas tilded quantities will always correspond to the
transformed scheme. For the case of the modification of the renormalisation
scheme, the same scale $s_0$ is assumed in both couplings $a_{s_0}$ and
$\tilde a_{s_0}$. In principle, one could have considered both transformations
simultaneously, in which case the coefficients $d_i$ would have corresponded
to coefficients $d_{i0}$ in eq.~\eqn{as0x}, but it appears more transparent to
discuss them separately.

Because the Adler function is a physical quantity, it should not depend on
the renormalisation scheme used. This can only be true, if together with the
coupling also the coefficients of the perturbative expansion $c_{n1}$ get
modified. Comparing in eq.~\eqn{Ds} or \eqn{Jna} the expansion in terms of the
two schemes, it is straightforward to read off the transformed coefficients
\cite{pen:82}. The scheme dependence of the coefficients only starts at order
$a^2$, which implies $\tilde c_{01} = c_{01}$ and $\tilde c_{11} = c_{11}$.
Up to order $a^5$ the remaining relations are found to be:
\begin{eqnarray}
\label{ctnk}
\tilde c_{21}  &=&  c_{21} - d_1 c_{11} \,, \quad
\tilde c_{31} \,=\, c_{31} - 2 d_1 c_{21} + ( 2 d_1^2 - d_2 )\,c_{11} \,,
\quad \nn \\
\smvs
\tilde c_{41}  &=&  c_{41} - 3 d_1 c_{31} + (5 d_1^2 - 2 d_2)\,c_{21} -
( 5 d_1^3 - 5 d_1 d_2 + d_3 )\,c_{11} \,, \\
\smvs
\tilde c_{51}  &=&  c_{51} - 4 d_1 c_{41} + ( 9 d_1^2 - 3 d_2  )\,c_{31} -
( 14 d_1^3 - 12 d_1 d_2 + 2 d_3 )\,c_{21} \nn \\
\smvs
&& \hspace{5.6mm} +\,( 14 d_1^4 - 21 d_1^2 d_2 + 3 d_2^2 + 6 d_1 d_3 - d_4 )
\,c_{11} \,. \nn
\end{eqnarray}
This shows that, even though CIPT still takes the functional form of
eq.~\eqn{RtauDCI}, now expressed in terms of the new coefficients
$\tilde c_{n1}$ as well as new contour integrals $\tilde J_n^a(s_0)$, assuming
the series to terminate in one scheme, this is no longer the case in a
different scheme. Through relations of the type \eqn{ctnk}, coefficients
$\tilde c_{n1}$ are generated to all orders. However, to be able to tell in
which scheme a certain coefficient $c_{n1}$ vanishes, it necessarily has to
be calculated in at least one reference scheme.

Like for eq.~\eqn{ctnk}, it is a simple matter to derive from \eqn{ats0} the
RGE for the coupling $\tilde a_{s_0}$ and the corresponding $\beta$-function
coefficients $\tilde\beta_i$. Thereby, one can confirm the well-known fact that
the first two coefficients of the $\beta$-function are independent of the
renormalisation scheme, i.e. $\tilde\beta_1=\beta_1$ and
$\tilde\beta_2=\beta_2$.\footnote{At least in renormalisation schemes which
are mass and gauge-parameter independent \cite{et:82}.} Starting with
$\tilde\beta_3$, it is found that the $\tilde\beta$ coefficients depend on the
$d_i$, such that the coefficient $\tilde\beta_k$ depends on all $d_i$ with
$i=1,\ldots,k-1$. Explicitly, the relations for $\tilde\beta_3$ and
$\tilde\beta_4$ take the form:
\begin{equation}
\label{betat}
\tilde\beta_3 \,=\, \beta_3 - d_1 \beta_2 - ( d_1^2 - d_2 ) \beta_1 \,, \quad
\tilde\beta_4 \,=\, \beta_4 - 2 d_1 \beta_3 + d_1^2 \beta_2 + ( 4 d_1^3 -
                    6 d_1 d_2 + 2 d_3 ) \beta_1 \,.
\end{equation}
Again, one could define a scheme, in which all $\tilde\beta_k = 0$ for
$k\geq 3$ \cite{tHo:77}. But in order to do this, one needs to know the
$\beta$-function coefficients in one reference scheme.

Let us now proceed with a numerical analysis of the scheme dependence. The
principle of minimal sensitivity (PMS) \cite{ste:81} attaches a special
importance to the point where the derivatives of the physical quantity with
respect to the scheme-parameters $d_i$ vanish, that is where the physical
quantity, with respect to the $d_i$ has an extremum or a saddle point and thus
the scheme dependence is weakest. Performing this exercise for
$R_\tau^{D,\rm{CI}}$ as a function of $d_1$ and $d_2$, and putting the
remaining $d_i$ to zero, one finds a maximum at $d_1 = 0.165$ and $d_2 = 0.210$.
(Trying to also include $d_3$ in the analysis, it is found that the extremum
drifts off to values of the $d$-coefficients, where the scheme transformation
\eqn{ats0} is no longer perturbative.) 

\FIGURE[t]{\includegraphics[angle=0, width=14cm]{Rtaud}
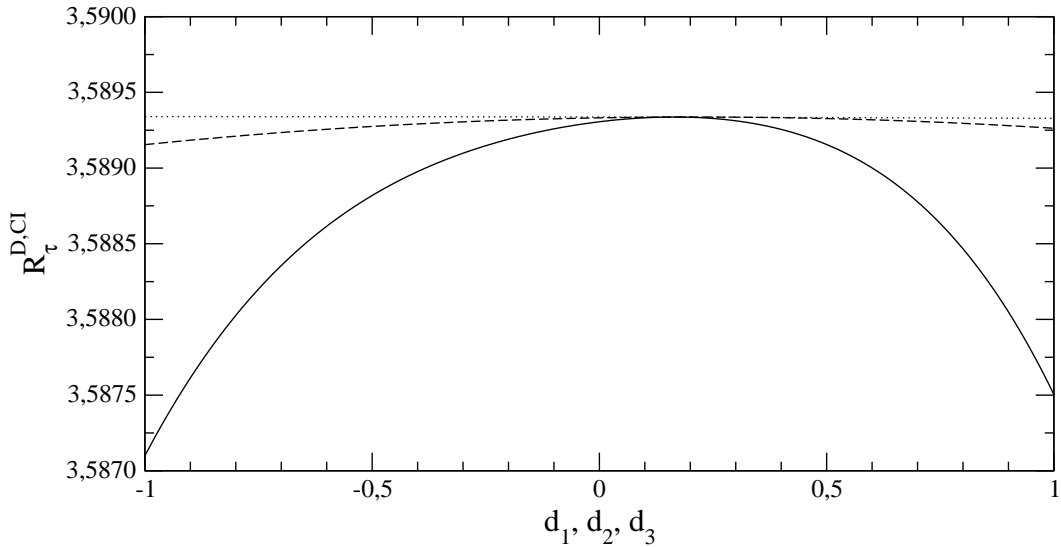
\caption{$R_\tau^{D,\rm{CI}}$ as a function of the renormalisation scheme
parameters $d_1$, $d_2$ and $d_3$. The solid line corresponds to a variation
of $d_1$, the dashed line to $d_2$ and the dotted line to $d_3$.
\label{fig5}}}

In figure~\ref{fig5}, the dependence of $R_\tau^{D,\rm{CI}}$, including up to
the fifth perturbative order, on the scheme parameters $d_1$, $d_2$ and $d_3$
is displayed in a graphical form. To this end, always only one parameter is
varied and the remaining parameters are kept fixed either at the maximum given
above for $d_1$ and $d_2$ or at zero for $d_3$ and $d_4$. In order that the
scheme transformation stays perturbative, a reasonable choice for the variation
of the $d_i$ seems to be the range $-1\leq d_i\leq 1$, such that the correction
is always at most about 10\% of the previous term. The variation of $d_1$ is
given by the solid line, with the maximum being at $d_1 = 0.165$. Next, the
variation of $d_2$ is given by the dashed line, with the maximum at
$d_2 = 0.210$, and finally the variation of $d_3$ corresponds to the dotted
line.

Several remarks are in order. As expected, for higher corrections, the
dependence on the scheme parameters $d_i$ gets weaker and weaker, being
strongest for $d_1$ and weakest for $d_3$. Actually, varying $d_3$, the change
in $R_\tau^{D,\rm{CI}}$ is only by one unit in the fifth digit. While varying
$d_1$, the maximal difference to the result of eq.~\eqn{RtauCIn} is given by
$3\cdot 0.00073$. Thus, the scheme variation is of a similar size as the
expected uncertainty from higher orders in the $\beta$-function, considered in
the last section. Furthermore, the ``optimal'' scheme, where the dependence on
the parameters $d_i$ is weakest lies rather close to the $\MSb$ scheme in
which all $d_i=0$. Therefore, also the speed of convergence of the
CI perturbative series in the $\MSb$ scheme can be expected to be close to
optimal.

\section{Conclusions}

Until today, the extraction of $\alpha_s$ from the hadronic $\tau$ decay rate
is hampered by an apparent ambiguity between performing the renormalisation
group improvement of the perturbative series at a fixed order or in the
so-called contour-improved scheme \cite{piv:91,dp:92a}. This ambiguity
represents the dominant theoretical uncertainty for $\alpha_s$ as extracted
from hadronic $\tau$ decays.

Further insight into the origin of the problem can be gained by pursuing the
fixed-order expansion to larger orders, still staying consistent with the
terms that have been resummed in contour-improved perturbation theory. In a
simplified example, where only the first coefficient of the $\beta$-function,
$\beta_1$, is kept non-vanishing, this could be done analytically in closed
form. For the more general case including the four currently known terms of
the $\beta$-function, the expansion of the fixed-order result can be calculated
to, in principle, arbitrarily high order. The behaviour of the fixed order
series in the simple example as well as in the more general case discussed in
sections~4 and 5 have been presented in tables~\ref{tab1} and \ref{tab2}
respectively, and graphically illustrated in figures~\ref{fig1} as well as
\ref{fig2}.

It is observed that in the full QCD case the fixed-order result oscillates
around the contour-improved value with an oscillation frequency of
approximately six perturbative orders, converging to it until the 30th order
after which an asymptotic behaviour of the series shows up and the terms again
increase. Just including five perturbative terms in the fixed-order case
happens to turn out most unfortunate, as for this order the deviation between
contour-improved and fixed-order results has a maximum. Analysing the
comparison of fixed-order as well as contour-improved perturbation theory for
an arbitrary renormalisation scale parameter $\xi$ unequal to one reveals that
also in this case both are compatible. However, invoking $\xi\neq 1$ in CIPT
reshuffles known contributions to higher orders thus throwing away available
information. Changing the renormalisation scheme is a somewhat different issue,
for there the independent coefficients $c_{n1}$ themselves get modified. Still,
this does not allow for an unambiguous assessment on the uncertainties resulting
from so far neglected higher orders.

All the discussion above demonstrates, that the renormalisation group
improvement of perturbative series in finite-energy sum rules, like for the
hadronic $\tau$ decay rate in particular, should be performed in the
contour-improved scheme. This approach also is most natural in the sense that
all terms proportional to a particular perturbative coefficient $c_{n1}$ in the
correlation function are resummed to yield the contour-improved contribution at
order $n$, whereas in the fixed-order approach terms proportional to the
coefficient $c_{n1}$ appear at all orders equal or greater to $n$. A discussion
of renormalisation group improvement of quark-mass corrections in finite-energy
sum rules along similar lines will be presented in a forthcoming publication.

\smallskip
\acknowledgments
It is a great pleasure to thank Martin Beneke, Andre Hoang, Andreas H\"ocker,
Santi Peris and Toni Pich for helpful discussions. This work has been supported
in part by the European Union RTN Network EURIDICE Grant No. HPRN-CT2002-00311,
and by MCYT (Spain) and FEDER (EU) Grants No. FPA2002-00748 and FPA-2001-3031.

\smallskip

\providecommand{\href}[2]{#2}\begingroup\raggedright\endgroup

\end{document}